\documentclass[booktabs,reprint,groupedaddress,amsmath,amssymb,aps,pra,floatfix]{revtex4-1}
\pdfoutput=1
\usepackage{amsmath,amssymb,graphicx,hyperref,mathrsfs,amsfonts,color,siunitx,setspace,xcolor,braket}

\newcommand{\SUPPLEMENTALHIDDEN}[1]{\iffalse\textcolor{red}{#1}\fi}

\newcommand{\OmegaTwoPh}{\Omega_{\text{ge}}}

\newcommand{\omegaHF}{\omega_{\text{HF}}}

\newcommand{\deltaAOM}{\delta_{\text{AOM}}} 
\newcommand{\SweepRange}{\Delta_{\text{swp}}}
\newcommand{\TwoPhotonDetuning}{\delta}

\newcommand{\Q}{Q}
\newcommand{\Pdiabatic}{P_{\text{d}}}
\newcommand{\Pscatter}{P_{\text{sc}}}
\newcommand{\Detuning}{\Delta}

\newcommand{\DecayRate}{\Gamma} 
\newcommand{\Rscat}{R_{\text{sc}}}
\newcommand{\INT}[4]{\int_{#1}^{#2} #3 \,\mathrm{d}#4}

\newcommand{\rhoii}{\rho_{\text{ii}}}
\newcommand{\rhoig}{\rho_{\text{ig}}}
\newcommand{\rhoie}{\rho_{\text{ie}}}
\newcommand{\rhogi}{\rho_{\text{gi}}}
\newcommand{\rhogg}{\rho_{\text{gg}}}
\newcommand{\rhoge}{\rho_{\text{ge}}}
\newcommand{\rhoei}{\rho_{\text{ei}}}
\newcommand{\rhoeg}{\rho_{\text{eg}}}
\newcommand{\rhoee}{\rho_{\text{ee}}}

\newcommand{\tilderhoig}{\tilde{\rho}_{\text{ig}}}
\newcommand{\tilderhoie}{\tilde{\rho}_{\text{ie}}}
\newcommand{\tilderhogi}{\tilde{\rho}_{\text{gi}}}

\newcommand{\tilderhoge}{\tilde{\rho}_{\text{ge}}}
\newcommand{\tilderhoei}{\tilde{\rho}_{\text{ei}}}
\newcommand{\tilderhoeg}{\tilde{\rho}_{\text{eg}}}

\newcommand{\dotrhoii}{\dot{\rho}_{\text{ii}}}
\newcommand{\dotrhoig}{\dot{\rho}_{\text{ig}}}
\newcommand{\dotrhoie}{\dot{\rho}_{\text{ie}}}

\newcommand{\dotrhogg}{\dot{\rho}_{\text{gg}}}

\newcommand{\dotrhoeg}{\dot{\rho}_{\text{eg}}}
\newcommand{\dotrhoee}{\dot{\rho}_{\text{ee}}}

\newcommand{\dottilderhoii}{\dot{\tilde{\rho}}_{\text{i}}}
\newcommand{\dottilderhoig}{\dot{\tilde{\rho}}_{\text{ig}}}
\newcommand{\dottilderhoie}{\dot{\tilde{\rho}}_{\text{ie}}}

\newcommand{\dottilderhoeg}{\dot{\tilde{\rho}}_{\text{eg}}}

\newcommand{\omegaa}{\omega_{\text{i}}}
\newcommand{\omegag}{\omega_{\text{g}}}
\newcommand{\omegae}{\omega_{\text{e}}}

\draft 

\hypersetup{
    colorlinks,
    linkcolor={red!70!black},
    citecolor={blue!50!black},
    urlcolor={blue!50!black}
}

\begin{document}

\title{Laser Cooling with Adiabatic Transfer on a Raman Transition} 



\author{Graham P. Greve}
\affiliation{JILA, NIST, and Department of Physics, University of Colorado, 440 UCB, 
Boulder, CO  80309, USA}
\author{Baochen Wu}
\affiliation{JILA, NIST, and Department of Physics, University of Colorado, 440 UCB, 
Boulder, CO  80309, USA}
\author{James K. Thompson}
\affiliation{JILA, NIST, and Department of Physics, University of Colorado, 440 UCB, 
Boulder, CO  80309, USA}
\email[]{graham.greve@colorado.edu}

\date{\today}

\begin{abstract}
A novel laser cooling mechanism was recently demonstrated using a narrow-linewidth optical transition in Ref.  \cite{ThompsonSrSweptCooling}. Counter-propagating laser beams are swept in frequency to cause adiabatic transfer between a ground state and excited state, and Doppler shifts provide time-ordering that ensures the associated photon recoils oppose the particle's motion. We now expand this technique to $^{87}$Rb, which has no narrow-linewidth optical transition, by using artificially-narrow two-photon Raman transitions. The cooling mechanism is capable of exerting large forces to compress the phase-space of the atomic ensemble without relying on spontaneous emission, providing further support for its potential use in cooling molecules or other particles that lack closed cycling transitions. Because the dynamics of the adiabatic transfer are crucial for assessing the feasibility of Raman SWAP cooling, we also develop a generic model for Raman Landau-Zener transitions in the presence of free-space scattering.
\end{abstract}


\maketitle 

Advances in laser cooling techniques have opened new scientific vistas for neutral atoms, ions, and mechanical resonators \cite{chu1986experimental,diedrich1989laser,chan2011laser}. Doppler cooling techniques have been widely applied to produce atoms and molecules at mK to sub-$\mu$K temperatures and high phase-space density \cite{PhysRevLett.110.263003, PhysRevA.84.061406, PhysRevA.61.061403, PhysRevLett.93.073003, vogel1999narrow, NarrowLineCoolingEr2013,NarrowLineCoolingTheory1989, NarrowLineCoolingSr1999, DoppCoolForbiddenTrans2000}.  However, standard Doppler cooling is limited in both final temperature and maximum force by the linewidth $\Gamma$ and wavelength $\lambda \equiv 2 \pi/k$ of the available optical transitions -- properties that are provided by nature and not under control of the experimentalist.  Further, extending Doppler cooling to molecules is difficult because after each photon absorption event, the molecule must undergo spontaneous decay to return to the ground state, but the molecule may be lost due to a large number of  ground states into which it can decay \cite{shuman2010laser, hummon20132d}.




Recently, a new laser-cooling mechanism called Sawtooth Wave Adiabatic Passage (SWAP) cooling was observed in $^{88}$Sr atoms \cite{ThompsonSrSweptCooling}. The forces that lead to cooling in SWAP cooling rely on adiabatic transitions back and forth between a ground state $\ket{g}= \ket{^1\mathrm{S}_0}$ and a long-lived optically excited state $\ket{e}=\ket{^3\mathrm{P}_1}$ with lifetime  $\tau = 1/\Gamma = 21~\mu$s.  This allows for the possibility of many photon-recoils being applied to the atom before spontaneous emission occurs, an important property for extending laser cooling to molecules.

\begin{figure}[!htb]
\includegraphics[width=3.375in]{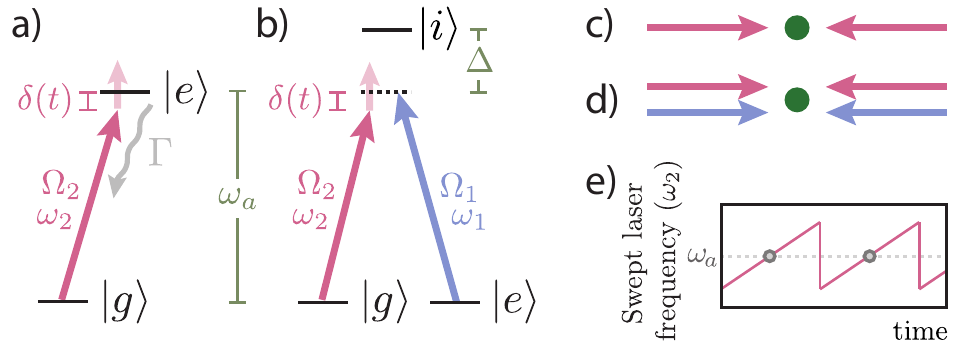}
\caption{SWAP cooling level diagram for \textbf{(a)} a single-photon narrow-linewidth transition and \textbf{(b)} a two-photon Raman transition. In the Raman scheme, the large detuning $\Delta$ from the intermediate state ensures the effective lifetime of the excited state $\ket{e}$ is effectively infinite except when it is useful to induce decay back to $\ket{g}$ following each sweep.  \textbf{(c)} In the single-photon case, the $\omega_2$ laser is counter-propagating and incident upon an atom (green). \textbf{(d)} The two-photon case is identical except a second $\omega_1$ laser serves to dress the $\ket{e}$ state. \textbf{(e)} The frequency of $\omega_2$ (red) is sawtooth-swept through the atomic transition frequency.}
\label{fig:comparison}
\end{figure}

In this paper, we present proof-of-principle experiments in $^{87}$Rb to demonstrate that the SWAP cooling mechanism can be extended to atoms and molecules with at least two long-lived ground states.  The core idea is that we will dress the ground states (here labeled $\ket{g}$ and $\ket{e}$) using  externally applied lasers tuned off-resonance from an intermediate optically excited state $\ket{i}$. This permits us to engineer effective optically excited states with tunable lifetimes or, equivalently, linewidths instead of relying on the properties of the optical transitions $\ket{g, e} \leftrightarrow \ket{i}$ (Fig.~\ref{fig:comparison}(a)).  A cooling scheme based on matter-wave interferometry has also been demonstrated with similar reported advantages \cite{InterferometricCooling}. Unlike the previous report of SWAP cooling with $^{88}$Sr, we achieve an equilibrium temperature in 1D that is 25 times lower than the usual $T_D\approx \frac{\hbar \Gamma}{2 k_B}= 146~\mu$K cooling limit for standard Doppler cooling in $^{87}$Rb.

In SWAP cooling using a single-photon transition, the magnetic field gradients, laser directions, and polarizations are essentially identical to those of standard Doppler cooling except that the pair of counter-propagating laser beams are ramped in frequency in a sawtooth pattern from below to above the transition frequency $\omega_a$.  The relative Doppler shift of the two laser beams causes the beam counter-propagating to the atom's motion to pass through resonance before the co-propagating beam.  As a result, the counter-propagating beam drives an adiabatic transition from $\ket{g}$ to $\ket{e}$ along with a momentum kick due to photon absorption that opposes the atomic motion.  The co-propagating beam then drives an adiabatic transition back from $\ket{e}$ to $\ket{g}$ along with a momentum kick due to stimulated emission that, again, slows the atom.  In net, each sweep ideally removes $2 \hbar k$ of momentum that reduces the atom's speed, regardless of the direction it is moving.  


In SWAP cooling using a two-photon transition, the frequency of the laser dressing the ground state $\ket{e}$ is held fixed.  The laser dressing the ground state $\ket{g}$ is swept in an asymmetric sawtooth pattern through a two-photon resonance, driving adiabatic two-photon Raman transitions between the ground states $\ket{g}$ and $\ket{e}$.  At the end of the sweep,  optical pumping is briefly applied to transfer atoms erroneously remaining in $\ket{e}$ back to $\ket{g}$.  In comparison to the work in strontium, this is equivalent to being able to set $\Gamma\approx0$ during the frequency sweep, but then setting $\Gamma$ up to the lifetime of the intermediate state for a very brief period of time in between sweeps, potentially offering a different degree of freedom for optimizing the cooling.

\begin{figure}[!htb]
\includegraphics[width=3.375in]{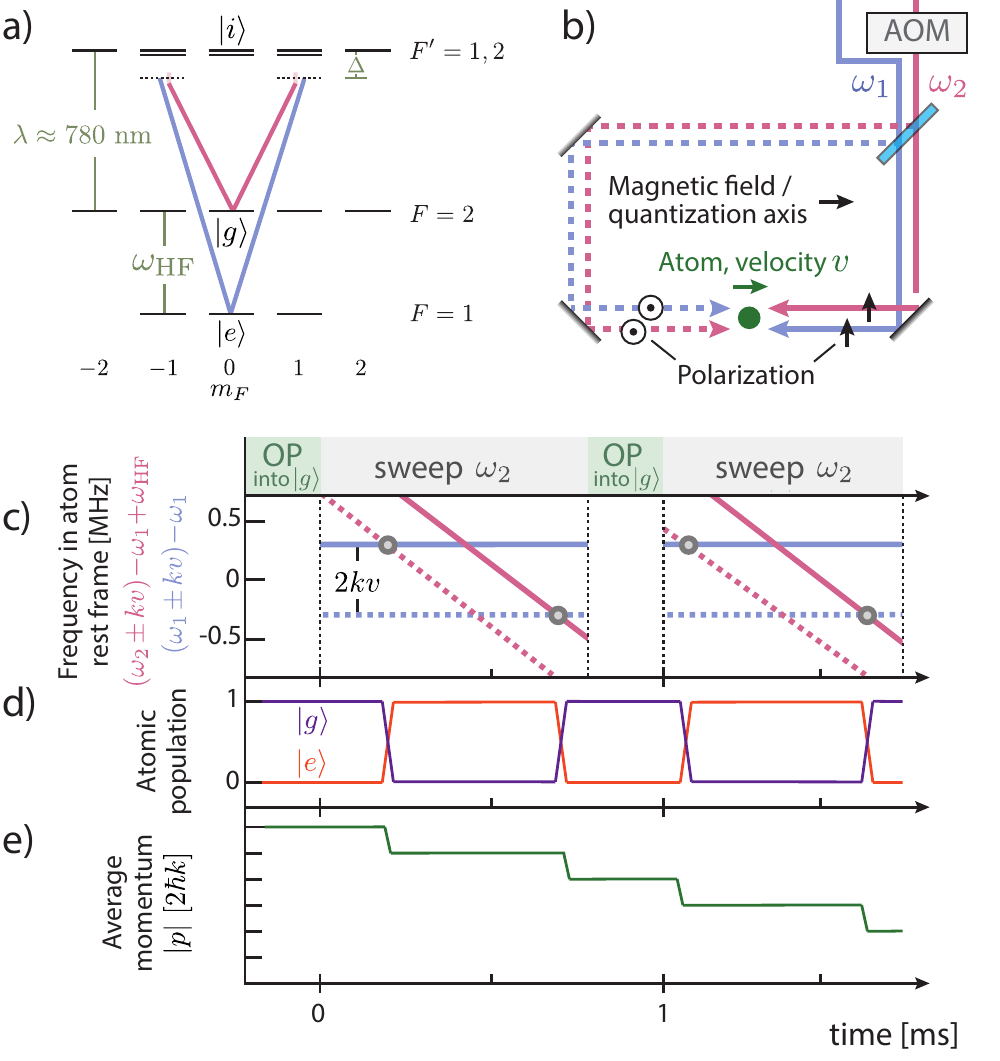}
\caption{\textbf{(a)} $^{87}$Rb level diagram. Two-photon transitions between $\ket{g}$ and $\ket{e}$ are induced by $\omega_1$ (blue) and $\omega_2$ (red). \textbf{(b)} Experimental layout. A moving atom (green) interacts with two pairs of counter-propagating laser beams (blue, red) with $\omega_2$ varied in a saw-tooth manner. \textbf{(c)} The four unique laser frequencies observed in the rest frame of the atom after including the Doppler shifts that separate counter-propagating lasers in frequency by $2 k v$. Offsets have been subtracted so that points marked with circles correspond to allowed two-photon resonances which involve pairs of orthogonally-polarized, counter-propagating laser beams. \textbf{(d)} Two-photon Landau-Zener transitions transfer an atom from $\ket{g}$ to $\ket{e}$ and back to $\ket{g}$ each sweep. \textbf{(e)} The counter-propagating lasers and sweep direction ideally remove $4 \hbar k$ momentum per cooling sweep.}
\label{fig:basics}
\end{figure}

Figure~\ref{fig:basics} shows the experimental setup for demonstrating 1D Raman SWAP cooling in $^{87}$Rb.  The quantization axis is established by a uniform magnetic field applied along the propagation axis of the cooling beams. The two ground states are the hyperfine Zeeman states $\ket{g}\equiv\ket{F=2, m_F=0}$ and $\ket{e}\equiv\ket{F=1, m_F=0}$. These states are coupled by lasers at frequencies $\omega_1$ and $\omega_2$, both far-detuned from the intermediate state $\ket{i}$ by an amount $\Delta\approx2 \pi \times 2$~GHz.  The allowed two-photon transitions involve absorption and stimulated emission of pairs of orthogonal linearly polarized photons differing in frequency by the ground state hyperfine splitting of $\omegaHF=6.834$~GHz. The cooling beams incident from the left are vertically polarized and the beams incident from the right are horizontally polarized.  Dipole selection rules disallow two-photon transitions with pairs of photons of the same linear polarization.  As a result, the only allowed two-photon Raman transitions are those that also impart a net photon recoil momentum $2 \hbar k$ as the internal state changes.



Both beam directions have two distinct frequency components $\omega_1$ and $\omega_2$  (Fig.~\ref{fig:basics}(b)) created by combining the output of two phase-locked lasers.  We choose to hold the frequency $\omega_1$ fixed while the frequency component $\omega_2$ is swept linearly in time downward in frequency through the two-photon atomic resonance at $\omega_1 -\omega_2 \approx \omegaHF$.  

Accounting for Doppler shifts of the laser frequencies as seen by an atom moving at speed $v$, there are two two-photon resonances that occur when $\delta(t) \equiv \omega_1-\omega_2-\omegaHF = \pm 2 k v$ (Fig.~\ref{fig:basics}(c)).  Ideal adiabatic passage through each resonance imparts a net momentum kick of $2 \hbar k$ per transition, \textit{i.e.} $4 \hbar k$ per sweep. The direction of the frequency sweep is chosen such that the time ordering of the passage through the two resonance frequencies leads to a reduction of the atom's speed in the laboratory frame.

As the atom approaches zero speed and the Doppler shift is of the same order as $\OmegaTwoPh$, the time-ordering of the adiabatic transfers detailed above is invalid. Here, $\OmegaTwoPh\approx \frac{\Omega_1 \Omega_2}{2\Delta}$ is the two-photon Rabi frequency and $\Omega_1\approx\Omega_2$ are the single photon Rabi frequencies of each frequency component. With the inevitable failure of a transfer, either due to time-ordering or imperfect adiabatic transfer, an atom can have considerable probability to be in $\ket{e}$ following a sweep so that the next iteration results in heating. We apply  $\pi$-polarized optical pumping light for $100~\mu$s to return any atoms remaining in $\ket{e}$ to $\ket{g}$ before the next frequency sweep.  




\begin{figure}[!htb]
\includegraphics[width=3.375in]{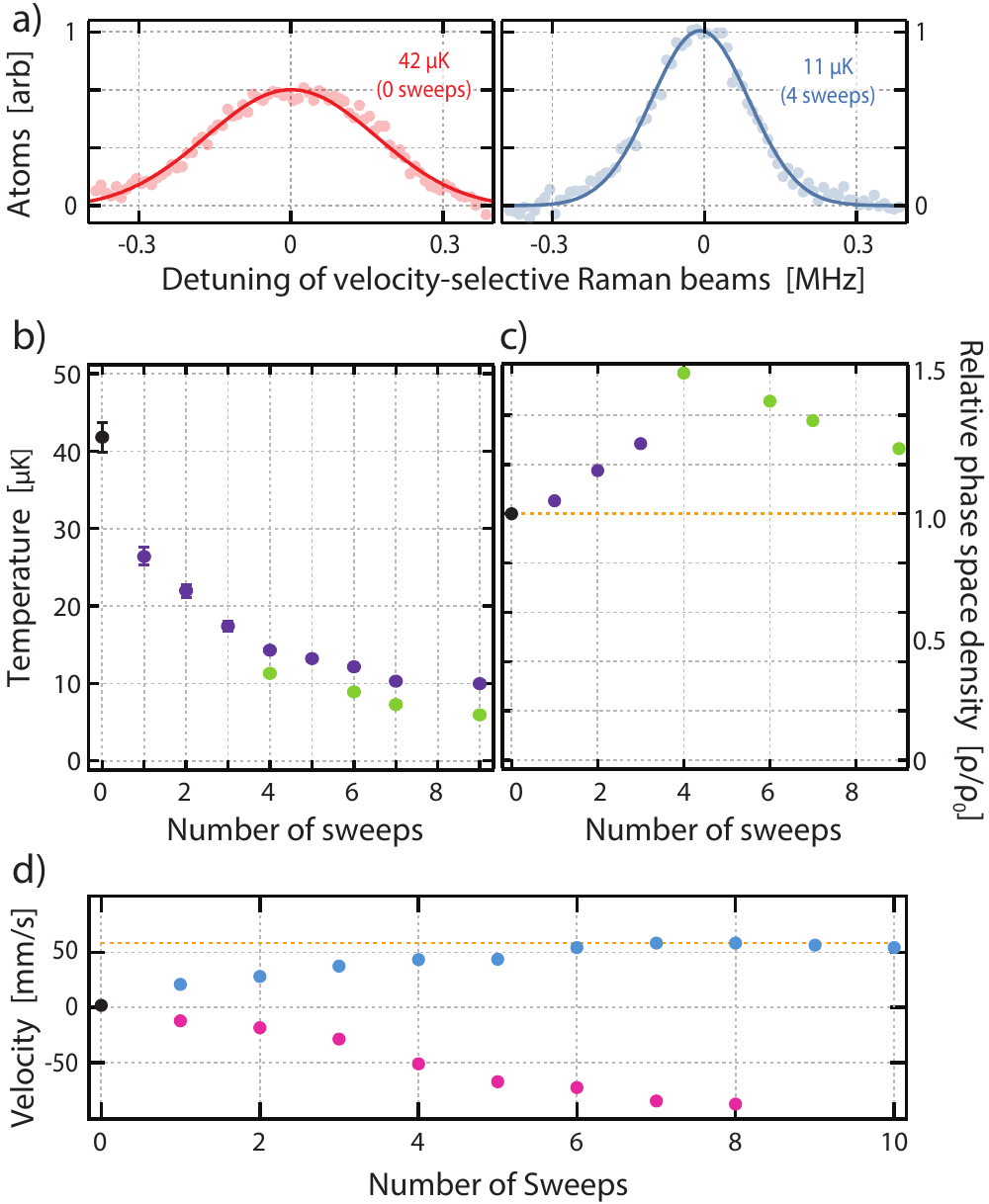}
\caption{\textbf{(a)} The reduction in temperature is determined by measuring the initial velocity distribution (red) and the distribution after four sweeps (blue). The velocity distributions shown here are determined by driving Raman transitions that are only resonant for atoms in a narrow velocity-class whose center is set by the detuning of the Raman beams.  \textbf{(b)} The atomic ensemble is cooled as low as 5.9 $\mu$K in one dimension over the course of several sweeps. For the purple points, all sweeps used a two-photon Rabi frequency $\OmegaTwoPh \approx 2\pi \times 22~\text{kHz}$ and sweep range $\SweepRange = 2 \pi \times 0.8~\text{MHz}$. For the green points, $\OmegaTwoPh$ and $\SweepRange$ were successively reduced between sweeps as described in the text. \textbf{(c)} The relative one-dimensional phase-space density for the corresponding purple or green points in (b). The phase-space density begins to decrease after four sweeps despite the decreasing temperature because atoms begin to leave the velocimetry beams. \textbf{(d)} To cool into a moving reference frame (blue points), we apply a frequency offset $\deltaAOM = 2 \pi \times 150~\text{kHz}$ in the lab reference frame between the beams from opposite directions. By sweeping $\omega_2$ downwards, the atoms cool and equilibrate into a moving reference frame which has a velocity $v_F \equiv \deltaAOM/2 k$ (dashed orange line).  If the sweep direction is reversed (pink points), the atoms accelerate in the opposite direction without being bound by $|v_F|$. 
}
\label{fig:cooling}
\end{figure}

In our experiment, around $10^7$ atoms are loaded into a magneto-optical trap (MOT) and pre-cooled with polarization gradient cooling (PGC) to narrow the initial velocity distribution and lessen the requirements on the sweep's effective capture range.  The atoms are optically pumped into $\ket{g}$. Frequency sweeps are then applied with typical sweep times of $1~$ms and sweep range $1~$MHz as shown in Fig.~\ref{fig:basics}(c).  The ideal internal state populations and change in momentum are shown in Fig.~\ref{fig:basics}(d and e).  This sequence of optical pumping and sweeping is repeated a number of times.  The final 1D temperature of the atoms is measured by velocimetry. All atoms are optically pumped back to $\ket{g}$ and then velocity-selective Raman transitions drive atoms within a small velocity range into $\ket{e}$ \cite{velocimetry}.  The population in $\ket{e}$ is determined using fluorescence and the resulting Voigt profiles are fit to extract the temperature (Fig.~\ref{fig:cooling}(a)).

We observe cooling from an initial temperature of $42(3)~\mu$K to $10~\mu$K after the application of 9 sweeps with fixed two-photon Rabi frequency $\OmegaTwoPh= 2 \pi \times 22$~kHz and sweep range of $\SweepRange \equiv 2 \pi \times \SI{0.8}{MHz}$  (red points in Fig.~\ref{fig:cooling}(b) ).  The sweep range is chosen to allow over 95\% of the atoms to pass through two-photon resonance during the sweep when accounting for the Doppler shifts of the initial velocity distribution of the atoms.

The temperature can be made as lower by progressively decreasing the two-photon Rabi frequency and the sweep range as is done for the green points in Fig.~\ref{fig:cooling}(b).  After the first three sweeps (blue points),  the Rabi frequency is reduced to $\sqrt{0.5} \times \OmegaTwoPh$ and the the sweep range is reduced to $0.5 \times \SweepRange$ such that the Landau-Zener adiabaticity parameter $\xi$ (discussed below) is unchanged.  The sweep range can be reduced because the velocity distribution was reduced by the initial sweeps.  After two sweeps, the remaining sweeps are performed with two-photon Rabi frequency  $\sqrt{0.4} \times \OmegaTwoPh$ and sweep range $0.4 \times \SweepRange$. After 9 total sweeps, the measured temperature reaches $5.9(3)~\mu$K.


 An increase in phase-space density demonstrates a reduction of entropy and not merely a selective loss of atoms or a redistribution of the density in phase space. In Fig.~\ref{fig:cooling}(c), we see that the the relative 1D phase space density $\rho/\rho_0 = \Delta x_0 \Delta v_0/(\Delta x \Delta v)$ is increased, where $\Delta x$ and $\Delta v$ ($\Delta x_0$ and $\Delta v_0$) are the measured cloud size and velocity spread after (before) cooling. 
 
 

Although the polarization scheme is reminiscent of polarization gradient cooling \cite{Dalibard:89}, here there is a large magnetic field present that breaks the degeneracy of the ground states that is typically required for polarization gradient cooling to such a low temperature. In addition, when the sweep direction was reversed, we observed heating as expected in the SWAP model of cooling.

To further emphasize the critical role of the sweep direction in the present work, we apply a fixed relative offset frequency $\deltaAOM$ between the counter-propagating beams such that one would expect that atoms are cooled into a moving reference frame with velocity $v_F = \deltaAOM/2k$.  If the laser frequency is swept downward as was done for the cooling experiments above, we observe that the atoms are accelerated into and equilibrate into the moving frame (blue points in Fig.~\ref{fig:cooling}(d).) The orange dashed line indicates the velocity of the predicted moving frame for the applied frequency offset $\deltaAOM = 2 \pi \times \SI{150}{\kHz}$.  In contrast, if we simply reverse the frequency sweep direction, we observe that the atoms are accelerated in the opposite direction despite the direction of the moving reference frame remaining unchanged (red points of Fig.~\ref{fig:cooling}(d). The atoms are accelerated to speeds larger than the calculated $|v_F|$ as expected.

It is important to understand the possible limitations of using Raman transitions associated with the sweep rate $\alpha\equiv \mathrm{d} \omega_2 /\mathrm{d}t$ being too fast or too slow.  To achieve high quality adiabatic transfer, one would like to sweep slowly such that the ideal Landau-Zener diabatic transition probability is small, $P_d= e^{-\xi}\ll1$ where ${\xi=\frac{\pi}{2} \OmegaTwoPh^2/\alpha}$.  To avoid off-resonant spontaneous scattering of light from the applied cooling laser beams, one would like the probability to have not scattered a photon during the total sweep time to be close to unity, $\Pscatter=e^{- \Rscat \SweepRange/\alpha}\approx 1$.  $\Rscat$ is the total spontaneous scattering rate from the far-from-resonance intermediate state.

\begin{figure}[!htb]
\includegraphics[width=3.375in]{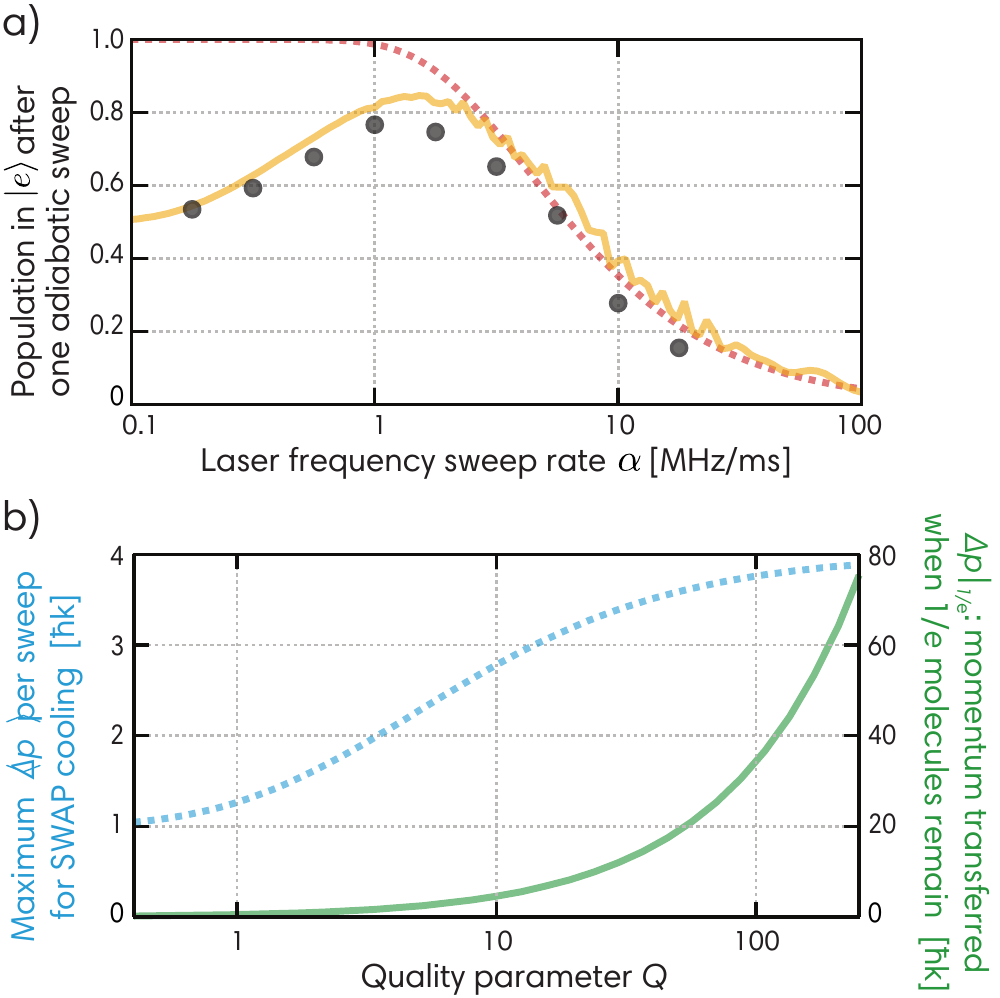}
\caption{\textbf{(a)} Population in $\ket{e}$ after one adiabatic transfer attempt, measured using fluorescence detection (black points). Here, transitions were driven using co-propagating $\sigma^{+}$-polarized light with frequencies $\omega_1$ and $\omega_2$. $\omega_1$ was constant and $\omega_2$ was swept through resonance at different sweep rates $\alpha = \mathrm{d} \omega_2 /\mathrm{d}t$.  Transfers were performed with $(\OmegaTwoPh, \Gamma, \Detuning, \SweepRange) = 2 \pi \times (\SI{20}{kHz}, \SI{6}{MHz}, \SI{2}{GHz}, \SI{0.8}{MHz})$. The familiar Landau-Zener prediction (red dashed) fails at low sweep rate due to off-resonant scattering from the intermediate state $\ket{i}$. A numerical simulation using the experimental parameters and including off-resonant scattering shows qualitative agreement (orange) with the data.  \textbf{(b, left)} The achievable momentum transfer per SWAP cooling sweep (blue dashed) approaches the ideal $4 \hbar k$ at large quality parameter $\Q \equiv \frac{\pi}{2} \frac{\Omega^2}{\Gamma \SweepRange}$.  \textbf{(b, right)} For SWAP cooling of molecules, we consider a worst-case scenario in which a molecule is lost if it undergoes a single spontaneous emission event. The predicted achievable momentum transfer when $1/e$ molecules remain (green) is shown versus the quality parameter $\Q$.}
\label{fig:theory}
\end{figure}

In Fig.~\ref{fig:theory}(a) (circles), we measure the probability to successfully transfer from $\ket{g}$ to $\ket{e}$ after a single adiabatic transfer using $\sigma^+$ polarized beams from a single direction and such that the two-photon Rabi frequency is the same as for the data in Fig.~\ref{fig:cooling}(b).  The data shows that there is an optimum sweep rate $\alpha$ that maximizes the transfer efficiency as desired for efficient cooling.  For comparison,  the red line is the predicted transfer efficiency $1-\Pdiabatic$ ignoring free space scattering.   The measurements qualitatively match predictions from numerically integrating optical-Bloch equations including spontaneous emission (Fig.~\ref{fig:theory}(a) orange).




To generalize predictions for how the tension between sweeping slowly to preserve adiabaticity and sweeping quickly to minimize scattering will limit SWAP cooling, we consider the three-level system shown in Fig.~\ref{fig:comparison}(b).   We will assume that the intermediate state $\ket{i}$ decays with equal rates of $\DecayRate/2$ into both $\ket{e}$ and $\ket{g}$. We also assume $\Omega_1=\Omega_2=\Omega$, and that the large detuning limit $\Delta\gg \DecayRate, \Omega$ is satisfied.  The two photon Rabi frequency is then $\OmegaTwoPh= \frac{\Omega^2}{2 \Delta}$, and the total scattering rate is $\Rscat = \frac{\Gamma \Omega^2}{4\Delta^2}$. More details about the following treatment are provided in the Supplemental Materials.


We use this idealized system to find the optimum momentum transferred $\Delta p$ during a cooling sweep after optimizing for sweep rate $\alpha$. In  Fig.~\ref{fig:theory}(b, blue), we plot $\Delta p$ against a dimensionless quality parameter ${\Q\equiv\ln{\Pdiabatic}/\ln{\Pscatter}} \approx\frac{\pi}{2}\frac{\Omega^2}{\Gamma \SweepRange}$. When $\Q \gg 1$, the momentum transfer per sweep saturates to the ideal value $4 \hbar k$, and the effective force is $4 \hbar k / t_{\text{swp}}$, where the time to complete each sweep is $t_{\text{swp}}$. At the optimized sweep rate, ${t_{\text{swp}} = \frac{8 \ln\left( 2 Q \right)}{\pi} \frac{\Delta^2}{\Omega^4} \SweepRange}.$ 
In terms of experimentally controllable parameters, the quality factor scales with laser intensity $I$ and wavelength as $\Q\propto I \lambda^3$, but does not depend on the dipole matrix element $M$ between the states $\ket{e}$, $\ket{g}$ and the intermediate state $\ket{i}$. The sweep time scales as roughly $t_{\text{swp}} \propto \frac{\Delta^2}{M^{4} I^{2}} \SweepRange$. 


For cooling molecules, where avoiding spontaneous emission is of chief importance, one must understand how much momentum can be removed before the molecule is lost. We take the worst case scenario, where every molecule that spontaneously emits a photon is completely lost. In  Fig.~\ref{fig:theory}(b, green), we plot $\Delta p |_{1/e}$, the average momentum transfer when $1/e$ molecules remain, again numerically optimizing the sweep rate. For $\Q \gg 1$, the numerical result is well approximated by ${\Delta p|_{1/e} \approx \frac{2.1 Q}{\ln\left(4\left(Q+14\right)\right)}  \hbar k}$. By engineering systems with high quality parameter $\Q$, one can remove many photon recoils of momentum from a molecule before it is likely to be lost. The optimized sweep time is nicely approximated by $t_{\text{swp}} \approx \frac{\ln\left( 4(14 + Q) \right)}{3.2} \frac{\Delta^2}{\Omega^4} \SweepRange$.


We have demonstrated that Raman transitions may be employed to produce SWAP cooling and achieve final temperatures well below the Doppler cooling limit. The technique is straightforward to implement, is amenable towards working in the presence of a large magnetic field,  is robust against small changes in atomic transition frequency, and might prove useful for cooling molecules. Future work may look towards using the technique to cool in more dimensions or to manipulate ensembles via accelerations and decelerations. More complex waveforms for the laser intensity and detuning could potentially decrease the required sweep range, increasing the effective cooling rate \cite{BatemanFreegarde}. We have also identified a quality parameter $\Q$ that provides guidance as to what experimental systems are needed for Raman SWAP cooling to work efficiently.

\section*{ACKNOWLEDGMENTS}\label{ack}
All authors acknowledge financial support from DARPA QuASAR, ARO, NSF PFC, and NIST. This work is supported by the National Science Foundation under Grant Number 1125844. 

\bibliographystyle{apsrev4-1}
\bibliography{main}

\clearpage 

\setcounter{equation}{0}
\setcounter{figure}{0}
\setcounter{table}{0}
\setcounter{section}{0}
\setcounter{page}{1}
\makeatletter
\renewcommand{\theequation}{A\arabic{equation}}
\renewcommand{\thefigure}{A\arabic{figure}}
\renewcommand{\thetable}{A\arabic{table}}
\renewcommand{\thesection}{A\Roman{section}}
\renewcommand{\thepage}{A\arabic{page}}

\begin{widetext}
\begin{spacing}{1.5}
\begin{center}\Large{\textbf{Supplementary Material}}\end{center}

As discussed in the main text, adiabatic transfer in the presence of scattering is optimized by a balance between the need to sweep slowly to preserve adiabaticity and the need to sweep fast enough to avoid significant scattering. We now present a derivation of the optical Bloch equations for a simplified three-level system to better understand the dynamics and limitations of adiabatic passage. Section \ref{app:a} includes the adiabatic elimination of the intermediate state $\ket{i}$. In Section \ref{app:b}, to distinguish successful adiabatic passage from a scattering event that transferred an atom into the correct final state, we account for the internal momentum state of the atom. In Section \ref{app:c}, results from the paper are explained in further detail and verified with the numerical models of the preceding sections.

We consider a slightly different model atom from the Main Text: this simplified model (Fig.~\ref{app:fig1}(a)) has stable ground states $\ket{g}$ and $\ket{e}$ and an optically excited state $\ket{i}$ that decays into them with equal probability at total rate $\Gamma$. The energy of these states are $\hbar \omegag$, $\hbar \omegae$, and $\hbar \omegaa$. A laser at frequency $\omega_1(t)$ couples $\ket{g} \leftrightarrow \ket{i}$ with single-photon Rabi frequency $\Omega_1$, and a laser at frequency $\omega_2(t)$ couples $\ket{e} \leftrightarrow \ket{i}$ with Rabi frequency $\Omega_2$. The average detuning $\Delta$ of these lasers from the excited state is large compared to the two-photon detuning $\TwoPhotonDetuning(t)$.

 \begin{figure}[!htb]\centering
\includegraphics[width=7.0in]{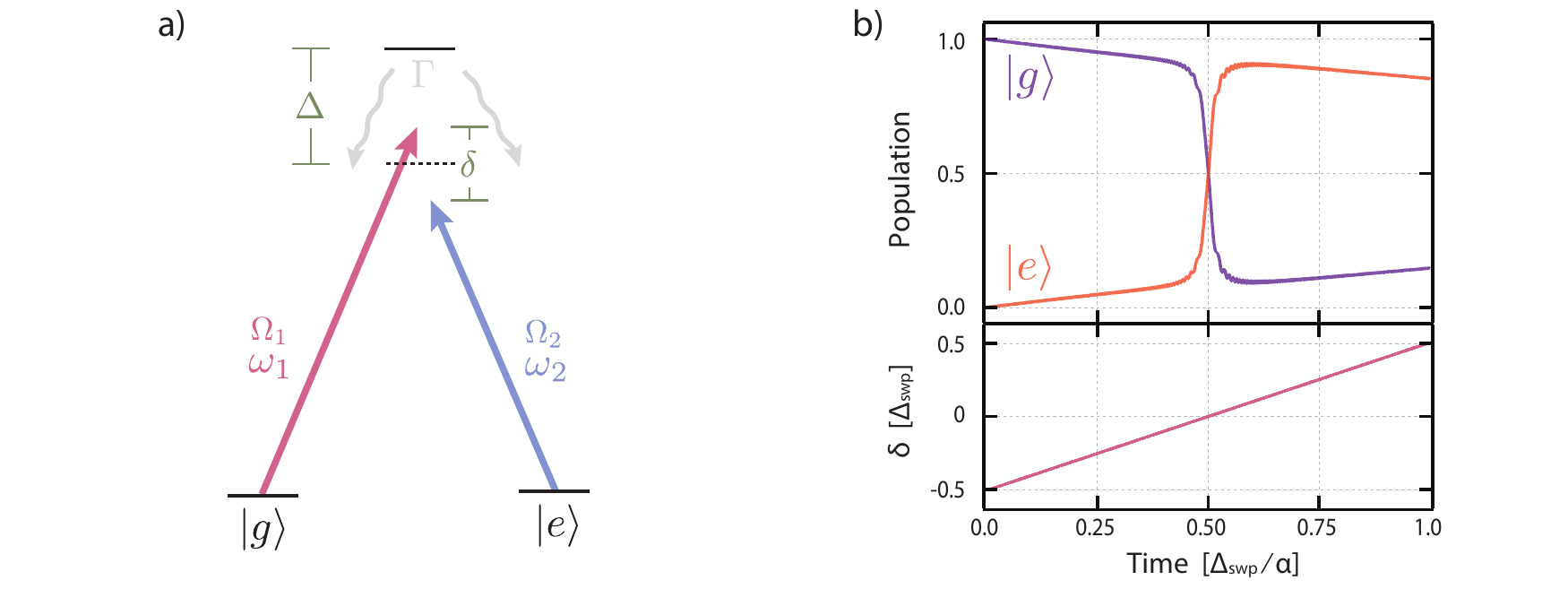}
\caption{\textbf{(a)} Diagram for simplified three-level system. An excited state $\ket{i}$ spontaneously decays equally to $\ket{g}$ and $\ket{e}$ at total decay rate $\Gamma$. Lasers connect $\ket{g}$ and $\ket{e}$ to a state detuned from $\ket{i}$ by $\Delta$. \textbf{(b)} Simulated time dynamics of adiabatic transfer. In units of the scattering rate, the sweep range is $\SweepRange = \Gamma/3$, $\Delta = 300 \Gamma$, $\OmegaTwoPh= \Gamma / 150$. The scattering rate $\Gamma$ is responsible for the coherence decay.}
\label{app:fig1}
\end{figure}

\section{Optical Bloch Equations and Adiabatic Elimination for a Three-Level Raman System}\label{app:a}

\noindent The density matrix describing this three-level system is \begin{align*}
    \rho &= \begin{pmatrix}
    \rhoii & \rhoig & \rhoie \\
    \rhogi & \rhogg & \rhoge \\
    \rhoei & \rhoeg & \rhoee
    \end{pmatrix}
\end{align*} and the Hamiltonian for this system can be written $\mathcal{H} = \mathcal{H}_A + V_1 + V_2$, with the interaction between an atom and a field provided by \begin{align*}
    V_{1} &= - \mathbf{d} \cdot \mathbf{E}_1 \cos\left(\omega_{1}(t) t\right) \\
    &= \frac{\hbar \Omega_{1}}{2} \left( \ket{i}\bra{g}\left(e^{-i \INT{0}{t}{\omega_1(t')}{t'}} + e^{i \INT{0}{t}{\omega_1(t')}{t'}} \right) + \ket{g}\bra{i}\left(e^{-i \INT{0}{t}{\omega_1(t')}{t'}} + e^{i \INT{0}{t}{\omega_1(t')}{t'}} \right) \right) \\
    &\approx \frac{\hbar \Omega_{1}}{2} \left( \ket{i}\bra{g} e^{i \INT{0}{t}{\omega_1(t')}{t'}} + \ket{g}\bra{i} e^{-i \INT{0}{t}{\omega_1(t')}{t'}} \right) \\
    \text{and} \qquad V_{2} &\approx \frac{\hbar \Omega_{2}}{2} \left( \ket{i}\bra{e}  e^{i \INT{0}{t}{\omega_2(t')}{t'}} + \ket{e}\bra{i} e^{-i \INT{0}{t}{\omega_2(t')}{t'}} \right)
\end{align*} after making the rotating wave approximation, with $\hbar \Omega_{1} \equiv - \Braket{g | \mathbf{d} \cdot \mathbf{E}_{1} | a}$ and $\hbar \Omega_{2} \equiv - \Braket{e | \mathbf{d} \cdot \mathbf{E}_{2} | a}$\SUPPLEMENTALHIDDEN{. Thus the Hamiltonian in matrix form is \begin{align*}
    \mathcal{H} &= \begin{pmatrix}
    \hbar \omegaa & \frac{\hbar \Omega_1 }{2} e^{-i \INT{0}{t}{\omega_1(t')}{t'}} & \frac{\hbar \Omega_2 }{2} e^{-i \INT{0}{t}{\omega_2(t')}{t'}} \\
    \frac{\hbar \Omega_1 }{2} e^{i \INT{0}{t}{\omega_1(t')}{t'}}  & \hbar \omegag & 0 \\
    \frac{\hbar \Omega_2 }{2} e^{i \INT{0}{t}{\omega_2(t')}{t'}}  & 0 & \hbar \omegae
    \end{pmatrix}
\end{align*}} where \begin{align*}
    \omega_1(t) &= \omegaa - \omegag  + \Delta(t) + \frac{\TwoPhotonDetuning(t)}{2} \\
    \omega_2(t) &= \omegaa - \omegae  + \Delta(t) - \frac{\TwoPhotonDetuning(t)}{2}
\end{align*} such that $\Delta(t)$ is the average detuning of the lasers from their respective transitions. In our experiment and simulations, $\TwoPhotonDetuning(t) = \alpha t - \frac{\SweepRange}{2}$ is swept linearly in time.

\noindent From the Liouville equation, $\dot{\rho} = \frac{i}{\hbar} \left[ \rho, \mathcal{H} \right] - \gamma \rho$, with the dissipation term $\gamma$ representing population relaxation, the equations of motion for the coherences are \begin{align}
    \dotrhoig &= -\left(i(\omegaa - \omegag) + \frac{\Gamma}{2}\right) \rhoig + \frac{i \Omega_1 e^{-i \INT{0}{t}{\omega_1(t')}{t'}}}{2} (\rhoii - \rhogg) - \frac{i \Omega_2 e^{-i \INT{0}{t}{\omega_2(t')}{t'}}}{2} \rhoeg \label{eq:xx}, \\
    \dotrhoie &= -\left(i(\omegaa - \omegae) + \frac{\Gamma}{2}\right) \rhoie + \frac{i \Omega_2 e^{-i \INT{0}{t}{\omega_2(t')}{t'}}}{2} (\rhoii - \rhoee) - \frac{i \Omega_1 e^{-i \INT{0}{t}{\omega_1(t')}{t'}}}{2} \rhoge, \\
    \dotrhoeg &= - i \left(\omegae - \omegag\right) \rhoeg + \frac{i \Omega_1 e^{-i \INT{0}{t}{\omega_1(t')}{t'}}}{2} \rhoei - \frac{i \Omega_2 e^{i \INT{0}{t}{\omega_2(t')}{t'}}}{2} \rhoig, \label{eq:xxxx}
\end{align} and for the populations, because the branching ratios from the intermediate state are balanced, \begin{align}
    \dotrhoii &= \frac{i \Omega_1}{2} \left( e^{i \INT{0}{t}{\omega_1(t')}{t'}} \rhoig - e^{-i \INT{0}{t}{\omega_1(t')}{t'}} \rhogi \right) + \frac{i \Omega_2}{2} \left( e^{i \INT{0}{t}{\omega_2(t')}{t'}} \rhoie - e^{-i \INT{0}{t}{\omega_2(t')}{t'}} \rhoei \right) - \Gamma \rhoii,  \label{eq:pop1} \\
    \dotrhogg &= \frac{i \Omega_1}{2} \left( e^{-i \INT{0}{t}{\omega_1(t')}{t'}} \rhogi - e^{i \INT{0}{t}{\omega_1(t')}{t'}} \rhoig \right)+ \frac{\Gamma}{2} \rhoii, \\
    \dotrhoee &= \frac{i \Omega_2}{2} \left( e^{-i \INT{0}{t}{\omega_2(t')}{t'}} \rhoei - e^{i \INT{0}{t}{\omega_2(t')}{t'}} \rhoie \right)+ \frac{\Gamma}{2} \rhoii. \label{eq:pop3}
\end{align}

\noindent To enter the rotating (natural) frame, the transformation to ``slow'' variables is used: \begin{align*}
    \rhoig &= \tilderhoig e^{-i \INT{0}{t}{\omega_1(t')}{t'}}, \\
    \rhoie &= \tilderhoie e^{-i \INT{0}{t}{\omega_2(t')}{t'}}, \\
    \rhoeg &= \tilderhoeg e^{-i \left(  \INT{0}{t}{\omega_1(t') - \omega_2(t')}{t'} \right)}
\end{align*} Substituting these into Eqns. \ref{eq:xx}-\ref{eq:xxxx}, the coherences in this frame become \begin{align*}
    \dottilderhoig &= \left[i \left( \Delta(t) + \frac{ \TwoPhotonDetuning(t)}{2}\right) - \frac{\Gamma}{2}\right] \tilderhoig + \frac{i \Omega_1}{2} (\rhoii - \rhogg) - \frac{i \Omega_2}{2} \tilderhoeg, \\
    \dottilderhoie &= \left[i \left( \Delta(t) - \frac{ \TwoPhotonDetuning(t)}{2} \right) - \frac{\Gamma}{2}\right] \tilderhoie + \frac{i \Omega_2}{2} (\rhoii - \rhoee) - \frac{i \Omega_1}{2} \tilderhoge, \\
    \dottilderhoeg &= \frac{i \Omega_1}{2} \tilderhoei - \frac{i \Omega_2}{2} \tilderhoig + i \TwoPhotonDetuning(t) \tilderhoeg.
\end{align*} The populations of Eqns. \ref{eq:pop1}-\ref{eq:pop3} can be rewritten \begin{align}
    \dotrhoii &= \frac{i \Omega_1}{2} \left( \tilderhoig - \tilderhogi \right) + \frac{i \Omega_2}{2} \left( \tilderhoie - \tilderhoei \right) - \Gamma \rhoii,  \label{eq:coh1} \\
    \dotrhogg &= \frac{i \Omega_1}{2} \left( \tilderhogi - \tilderhoig \right)+ \frac{\Gamma}{2} \rhoii, \label{eq:coh2} \\
    \dotrhoee &= \frac{i \Omega_2}{2} \left( \tilderhoei - \tilderhoie \right)+ \frac{\Gamma}{2} \rhoii. \label{eq:coh3}
\end{align}

We now reduce the equations of motion to an effective two-level system by adiabatically eliminating $\rhoig$ and $\rhoie$. This is justified when the population of the intermediate state is small, which is valid when $\Delta \gg \Gamma, \TwoPhotonDetuning$. The time derivatives $\dottilderhoig,$  $\dottilderhoii,$ and $\dottilderhoie$ are set to zero, and hence \begin{align*}
    \tilderhoig &= \frac{i \Omega_1}{\Gamma + 2 i \Delta(t) - i \TwoPhotonDetuning(t)} \left( \rhoii - \rhogg \right) - \frac{i \Omega_2}{\Gamma + 2 i \Delta(t) - i \TwoPhotonDetuning(t)} \tilderhoeg, \\
    \tilderhoie &= \frac{\Omega_2}{\Gamma + 2 i \Delta(t) + i \TwoPhotonDetuning(t)} \left( \rhoii - \rhoee \right) - \frac{\Omega_1}{\Gamma + 2 i \Delta(t) + i \TwoPhotonDetuning(t)} \tilderhoge.
\end{align*}

These coherences are substituted into Eqns.~\ref{eq:coh1}-\ref{eq:coh3}, and after transforming back into the original coordinates, we arrive at the final set of Bloch equations \begin{align*}
\dotrhogg &= \frac{\Gamma \rhoii}{2} + \frac{\Gamma \Omega_1^2  \left(\rhoii - \rhogg\right) }{\Gamma^2 + (2 \Delta(t) + \TwoPhotonDetuning(t))^2} - \frac{\Omega_1 \Omega_2}{2} \left( \frac{ \tilderhoeg}{\Gamma - 2 i \Delta(t) - i \TwoPhotonDetuning(t)} + \frac{\tilderhoge}{\Gamma + 2 i \Delta(t) + i \TwoPhotonDetuning(t)} \right), \\
\dotrhoee &= \frac{\Gamma \rhoii}{2} + \frac{\Gamma \Omega_2^2  \left(\rhoii - \rhoee\right) }{\Gamma^2 + (2 \Delta(t) - \TwoPhotonDetuning(t))^2} - \frac{\Omega_1 \Omega_2}{2} \left( \frac{\tilderhoeg}{\Gamma + 2 i \Delta(t) - i \TwoPhotonDetuning(t)} + \frac{\tilderhoge}{\Gamma - 2 i \Delta(t) + i \TwoPhotonDetuning(t)} \right),\\
    \dottilderhoeg &= i \TwoPhotonDetuning(t) \tilderhoeg - \frac{\tilderhoeg}{2} \left( \frac{\Omega_1^2}{\Gamma + 2 i \Delta(t) - i \TwoPhotonDetuning(t)}  + \frac{\Omega_2^2}{\Gamma - 2 i \Delta(t) - i \TwoPhotonDetuning(t)} \right) - \frac{\Omega_1 \Omega_2}{2} \left( \frac{\rhoee - \rhoii}{\Gamma + 2 i \Delta(t) - i \TwoPhotonDetuning(t)} + \frac{\rhogg - \rhoii}{\Gamma - 2 i \Delta(t) - i \TwoPhotonDetuning(t)} \right)\end{align*} with \begin{align}
    \rhoii &\approx \frac{ \Omega_1^2 \rhogg + \Omega_2^2 \rhoee + \Omega_1 \Omega_2 \tilderhoeg + \Omega_1 \Omega_2 \tilderhoge }{\Gamma^2 + 4 \Delta(t)^2 + \Omega_1^2 + \Omega_2^2}. \label{eq:rhoiifinal}
\end{align}

Equation \ref{eq:rhoiifinal} has been greatly simplified by the assumption that $\TwoPhotonDetuning(t) \ll \Delta(t), \Omega_1, \Omega_2$. \SUPPLEMENTALHIDDEN{The $\rhoii$ population is small, but it plays an important role in the system dynamics. It has now been rewritten in terms of the ground and excited state elements, and the system is thus reduced to a two-level density matrix \begin{align*}
    \rho_2(t) &= \begin{pmatrix}\rhogg & \rhoge \\ \rhoeg & \rhoee \end{pmatrix}.
\end{align*} }

\section{Including Internal Momentum States}\label{app:b}
 We now include the momentum states of an atom, using a second label such as $\ket{g, 0 \hbar k}$. The basis of states is truncated to only include $0\hbar k, 1 \hbar k, 2 \hbar k, \dots, (m-1) \hbar k$  momentum states, so the density matrix becomes $2m \times 2m$. Terms that are non-physical, involving a momentum transfer with no change of state, can also be negated:
 
\begin{align*}
\rho = \begin{pmatrix}
\ddots & \vdots & & \vdots &\ddots & \vdots & & \vdots & \\
\dots & \rho_{\text{gg00}} & 0 & 0 & \dots & \rho_{\text{ge00}} & \rho_{\text{ge01}} & \rho_{\text{ge02}} & \dots \\
& 0 & \rho_{\text{gg11}} & 0 & & \rho_{\text{ge10}} & \rho_{\text{ge11}} & \rho_{\text{ge12}} & \\
\dots & 0 & 0 & \rho_{\text{gg22}} & \dots & \rho_{\text{ge20}} & \rho_{\text{ge21}} & \rho_{\text{ge22}} & \dots \\
\ddots & \vdots & & \vdots & \ddots & \vdots & & \vdots & \ddots \\
\dots & \rho_{\text{eg00}} & \rho_{\text{eg01}} & \rho_{\text{eg02}} & \dots & \rho_{\text{ee00}} & 0 & 0 & \dots \\
& \rho_{\text{eg10}} & \rho_{\text{eg11}} & \rho_{\text{eg12}} & & 0 & \rho_{\text{ee11}} & 0 & \\
\dots & \rho_{\text{eg20}} & \rho_{\text{eg21}} & \rho_{\text{eg22}} & \dots & 0 & 0 & \rho_{\text{ee22}} & \dots \\
& \vdots && \vdots & \ddots & \vdots && \vdots & \ddots
\end{pmatrix}
\end{align*}

Normal adiabatic passage from $\ket{g, 0 \hbar k}$ through $\ket{i, 1 \hbar k}$ to $\ket{e, 2 \hbar k}$ is represented in the matrix elements
\begin{align*}
\dot{\rho}_{\text{gg00}} &= \frac{\Gamma \rho_{\text{ii00}}}{2} + \frac{\Gamma \Omega_1^2  \left(\rho_{\text{ii11}} - \rho_{\text{gg00}}\right) }{\Gamma^2 + (2 \Delta(t) + \TwoPhotonDetuning(t))^2} - \frac{\Omega_1 \Omega_2}{2} \left( \frac{ \tilde{\rho}_{\text{eg20}}}{\Gamma - 2 i \Delta(t) - i \TwoPhotonDetuning(t)} + \frac{\tilde{\rho}_{\text{ge02}}}{\Gamma + 2 i \Delta(t) + i \TwoPhotonDetuning(t)} \right), \\
\dot{\rho}_{\text{ee22}} &= \frac{\Gamma \rho_{\text{ii22}}}{2} + \frac{\Gamma \Omega_2^2  \left(\rho_{\text{ii11}} - \rho_{\text{ee22}}\right) }{\Gamma^2 + (2 \Delta(t) - \TwoPhotonDetuning(t))^2} - \frac{\Omega_1 \Omega_2}{2} \left( \frac{\tilde{\rho}_{\text{eg20}}}{\Gamma + 2 i \Delta(t) - i \TwoPhotonDetuning(t)} + \frac{\tilde{\rho}_{\text{ge02}}}{\Gamma - 2 i \Delta(t) + i \TwoPhotonDetuning(t)} \right), \\
    \dot{\tilde{\rho}}_{\text{eg20}} &= \frac{\tilde{\rho}_{\text{eg20}}}{2} \left( 2 i \TwoPhotonDetuning(t) - \frac{\Omega_1^2}{\Gamma + 2 i \Delta(t) - i \TwoPhotonDetuning(t)}  - \frac{\Omega_2^2}{\Gamma - 2 i \Delta(t) - i \TwoPhotonDetuning(t)} \right) \nonumber \\
    & \qquad \qquad \qquad \qquad - \frac{\Omega_1 \Omega_2}{2} \left( \frac{\rho_{\text{ee22}} - \rho_{\text{ii11}}}{\Gamma + 2 i \Delta(t) - i \TwoPhotonDetuning(t)} + \frac{\rho_{\text{gg00}} - \rho_{\text{ii11}}}{\Gamma - 2 i \Delta(t) - i \TwoPhotonDetuning(t)} \right)
\end{align*} with \begin{align*}
    \rho_{\text{ii11}} &\approx \frac{ \Omega_1^2 \rho_{\text{gg00}} + \Omega_2^2 \rho_{\text{ee22}} + \Omega_1 \Omega_2 \tilde{\rho}_{\text{eg20}} + \Omega_1 \Omega_2 \tilde{\rho}_{\text{ge02}} }{\Gamma^2 + 4 \Delta(t)^2 + \Omega_1^2 + \Omega_2^2}.
\end{align*}

Other momentum states are coupled in due to the scattering terms. For example, $\ket{e, 2 \hbar k}$ may absorb a photon, losing $\hbar k$ of momenta, and enter $\ket{a, 1 \hbar k}$. This state may in turn decay incoherently to $\ket{g, 1 \hbar k}$ or $\ket{e, 1 \hbar k}$.

Because the momentum states are now distinguishable within the density matrix, we can separate atoms that undergo adiabatic passage without scattering from other possibilities (Fig.~\ref{app:fig2}). We can also construct expectation values for the momentum change.

 \begin{figure}[!htb]\centering
\includegraphics[width=6.0in]{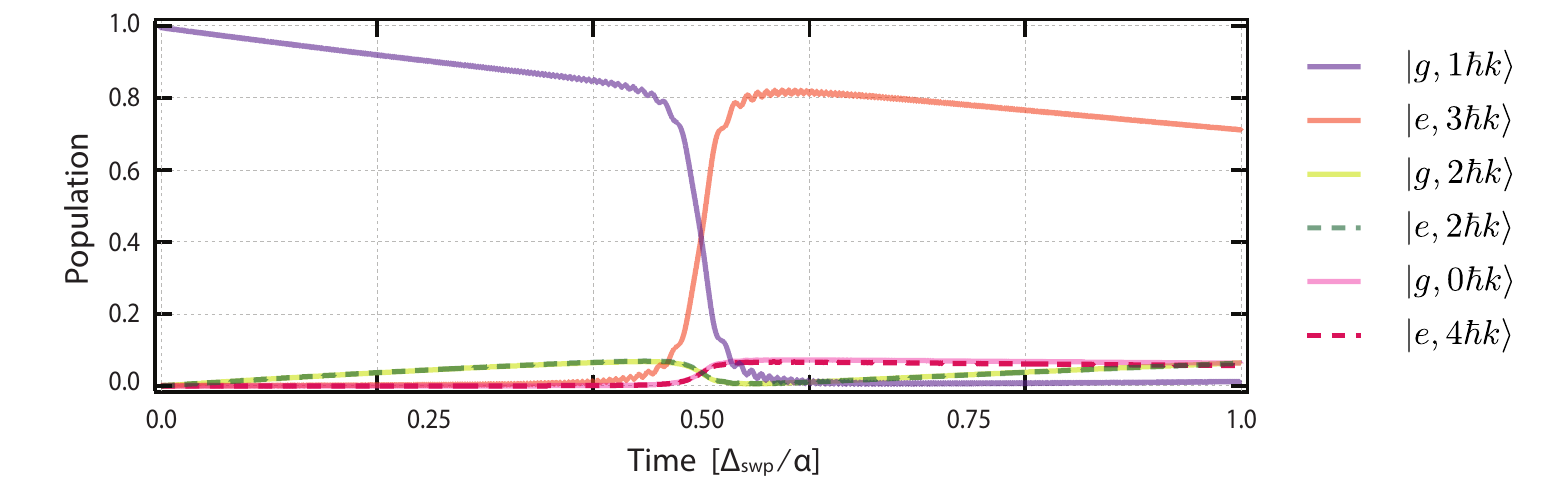}
\caption{\textbf{(a)} Simulation of populations including momentum labels during adiabatic transfer. The parameters used match those of Fig.~\ref{app:fig1}(b). The expected change in momentum at the end of the adiabatic transfer is $1.71 \hbar k$.}
\label{app:fig2}
\end{figure}

A Runge-Kutta fourth-order numerical integration method is used to simulate adiabatic transfer (Fig. \ref{app:fig1}(b)).  In practice, we find a $10\times10$ density matrix is sufficient, and population starts in $\ket{g, 1\hbar k}$ ($\rho_{\text{gg11}}$).  When determining final populations, oscillations in the simulation can become significant, so we take the average of  during the last 1/30th of time steps. This leads to a slight error due to the average slope during this time. To find the expected change of momentum, we compute \begin{align*}
   \Braket{ \Delta p(\Gamma, \Delta, \Omega, \SweepRange)_{\text{AT}}} &= \left[ \sum_{n} n\, \hbar k \left( \rho_{eenn} +  \rho_{ggnn} \right) \right] - 1 \hbar k.
\end{align*}

Finally, this model ignores effects such as laser noise and atomic dephasing which can set substantial limitations on transfer fidelity \cite{Lacour, AdiabaticPassageWithNoise}. In the main text's Fig.~\ref{fig:theory}(a), a separate two-level simulation was used which allowed us to enter this dephasing by hand. However, these details are beyond the scope of this model.

\section{A Probabilitistic Model for Raman SWAP Cooling}\label{app:c}

Adiabatic transfer from $\ket{g}$ to $\ket{e}$ adds $2 \hbar k$ momentum from an atom, and transfer from $\ket{e}$ to $\ket{g}$ removes $2 \hbar k$ momentum. If we assume that during a scattering process, absorption from $\ket{g}$ adds one photon's worth of momentum, and absorption from $\ket{e}$ removes one photon's worth momentum, but decay into either state causes no coherent momentum change, then to first-order, the possible momentum changes are detailed in Table \ref{tab:transfer_possibilities}. The first-order expected change in momentum, from summing the tabulated momentum changes multiplied by the probability of that trajectory, is found to be \begin{align*}
    \left\langle \Delta p_{\text{AT}} \right\rangle &= \left( 1 + x - 2 x^{1 + \Q} \right) \times \hbar k \label{eq:expmom}
\end{align*}  with $x \equiv e^{- \Rscat \frac{\SweepRange}{\alpha}}$, $\Rscat = \frac{\Gamma \Omega^2}{\Gamma^2 + 4 \Delta^2 + 2 \Omega^2} \approx \frac{\Gamma \Omega^2}{4 \Delta^2}$, and $\Q \equiv \frac{\pi}{2}\frac{\Omega^2}{\Gamma \SweepRange}$. The simulations of the previous section serve to validate this model exceptionally well.

\begin{table}
    \centering
    \begin{tabular}{c|l|l|l|c|c}
        \textbf{Row} & \textbf{Before $T/2$} & \textbf{At $T/2$} & \textbf{After $T/2$} & \textbf{Probability} & \textbf{Final $\Delta p$} \\\hline\hline
        1 & Scatter to $\ket{e}$ & Adiabatic transfer to $\ket{g}$ & Scatter & $\frac{1}{2} SAS$ & $0\, \hbar k$ \\
        2 & Scatter to $\ket{e}$ & Adiabatic transfer to $\ket{g}$ & No scatter & $\frac{1}{2} SA(1-S)$ & $-1\, \hbar k$ \\
        3 & Scatter to $\ket{e}$ & No adiabatic transfer & Scatter & $\frac{1}{2} S(1-A)S$ & $0\, \hbar k$ \\
        4 & Scatter to $\ket{e}$ & No adiabatic transfer & No scatter & $\frac{1}{2} S(1-A)(1-S)$ & $1\, \hbar k$ \\
        5 & Scatter to $\ket{g}$ & Adiabatic transfer to $\ket{e}$ & Scatter & $\frac{1}{2} SAS$ & $2\, \hbar k$ \\
        6 & Scatter to $\ket{g}$ & Adiabatic transfer to $\ket{e}$ & No scatter & $\frac{1}{2} SA(1-S)$ & $3\, \hbar k$ \\
        7 & Scatter to $\ket{g}$ & No adiabatic transfer & Scatter & $\frac{1}{2} S(1-A)S$ & $2\, \hbar k$ \\
        8 & Scatter to $\ket{g}$ & No adiabatic transfer & No scatter & $\frac{1}{2} S(1-A)(1-S)$ & $1\, \hbar k$ \\
        9 & No scatter & Adiabatic transfer to $\ket{e}$ & Scatter & $(1-S)AS$ & $1\, \hbar k$ \\
        10 & No scatter & Adiabatic transfer to $\ket{e}$  & No scatter & $(1-S)A(1-S)$ & $2\, \hbar k$ \\
        11 & No scatter & No adiabatic transfer & Scatter & $(1-S)(1-A)S$ & $1\, \hbar k$ \\
        12 & No scatter & No adiabatic transfer & No scatter & $(1-S)(1-A)(1-S)$ & $0\, \hbar k$ \\
    \end{tabular}
    \caption{First-order possibilities for atom state trajectories. An atom starts in $\ket{g}$ and may scatter before time $T/2$. At time $T/2$, adiabatic transfer may occur. From $T/2$ to $T$, the atom may scatter again. The probability to undergo adiabatic transfer is $A \equiv 1 - \exp\left(- \frac{\pi}{2} \frac{\OmegaTwoPh^2}{\alpha} \right)$, and the probability to scatter is $S \equiv 1 - \exp\left(- \Rscat \frac{\SweepRange}{2 \alpha}  \right)$ with scattering rate $\Rscat = \frac{\Gamma \Omega^2}{\Gamma^2 + 4 \Delta^2 + 2 \Omega^2} \approx \frac{\Gamma \Omega^2}{4 \Delta^2}$.}
    \label{tab:transfer_possibilities}
\end{table}

The momentum transfer during a full SWAP cooling sweep comes from cooling during the first half of the sweep, and cooling or heating during the second half depending on if an atom successfully transferred from $\ket{g}$ to $\ket{e}$. The expected momentum transfer during a full cooling sweep is then  $\Braket{\Delta p_{\text{SWAP}}} = \Braket{\Delta p_{\text{AT}}} \left( 1 + \left(P(e) - P(g)\right) \right)$. The fraction of atoms cooled vs. heated in the second half of the sweep is represented by $\left( P(e) - P(g)\right)$, where $P(a)$ is the probability to be in state $\ket{a}$ at the conclusion of the first half of the sweep. We find  \begin{equation}
    \Braket{\Delta p_{\text{SWAP}}} = \left\langle \Delta p_{\text{AT}} \right\rangle \left( 1 + x (1 - 2x^Q) \right).
    \label{eq:expMomSweep}
\end{equation}

To evaluate the benefits of SWAP cooling over Doppler cooling, we would like to know how much momentum could be removed before an atom (or molecule) detrimentally scatters a photon. Although we assumed equal branching ratios up until now, suppose that every scattering event causes an atom to be lost. We assume each atom scatters one photon from optical repumping if it finishes the cooling sweep in $\ket{e}$. The expected change in momentum per sweep comes entirely from the events of Table~\ref{tab:transfer_possibilities} Row 10 and 12: \begin{align*}
    \Braket{\Delta p_{\text{AT}}'} &= (1-S)A(1-S) \times 2 \hbar k = x(1-x^Q) \times 2 \hbar k
    \end{align*}

To be consistent, we need to normalize this to the number of atoms remaining in the system: \begin{align*}
    \Rightarrow \qquad \Braket{\Delta p_{\text{AT}}} &= \frac{ \Braket{\Delta p_{\text{AT}}'} }{ (1-S)A(1-S) + (1-S)(1-A)(1-S) } \\
    &= (1-x^Q) \times 2 \hbar k \\
    \Rightarrow \qquad 
    \Braket{\Delta p_{\text{SWAP}}} &= \Braket{\Delta p_{\text{AT}}}  \times \left[ 1 + \left( (1-S)A(1-S) - (1-S)(1-A)(1-S) \right) \right] \\
    &= \left( 1 + x - x^{Q} - 3x^{1+Q} + 2 x^{1+2Q} \right) \times 2 \hbar k
\end{align*}

Next we choose to find the average momentum transferred per atom at a time when $1/e$ atoms remain, with the assumption that any scattering event will cause the atom to be lost. The two ways an atom does not scatter a photon (during the sweep or from optical repumping) come from Rows 10 and 12: \begin{align*}
    P(\text{no scatter}) &= \left[ (1-S)A(1-S) \right]^2 + \left[ (1-S)(1-A)(1-S) \right]^2.
\end{align*} Let the probability that an atom survives $n$ sweeps be $1/e$; then the momentum change for an atom at the point when it has probability $1/e$ not to be lost is \begin{align}
     n \Braket{\Delta p_{\text{SWAP}}} &= \frac{ \Braket{\Delta p_{\text{SWAP}}} }{ - \ln\left( P(\text{no scatter}) \right) } \nonumber\\
    &= \left[ \frac{ (1-x^Q)(x(2x^Q - 1) - 1)}{\ln\left( x^2+2x^{Q+2}(x^Q-1) \right) } \right] \times 2 \hbar k.
    \label{eq:momentumBefore1OverE}
\end{align} This function is plotted in Fig.~\ref{fig:momTransfer}(b) for $Q=25$ and maximized with respect to sweep rate in Fig.~\ref{fig:theory}(b) in the Main Text.

The optimal sweep rate for transferring momentum in the model where scattering is unimportant is found by taking the derivative of Eqn.~\ref{eq:expMomSweep} with respect to $\alpha$ and equating it to zero: \begin{align}\label{eq:alphaopt}
    \alpha_{\text{opt}} &\approx \frac{\pi}{2} \frac{\Omega^4}{4 \Delta^2} \frac{1}{\ln\left[ 2\left( 1 + \Q \right) \right]}. 
\end{align} However, the optimal sweep rate for transferring momentum before atoms are lost to recoiled photons (Eqn. \ref{eq:momentumBefore1OverE}) could only be found numerically (Fig.~\ref{fig:momTransfer}(a)). The scaling at significant $Q$ remains the same as in Eqn.~\ref{eq:alphaopt} -- the optimal sweep rate can be increased as roughly $\Omega^4 / \Delta^2$ (though $\Q$ also changes as $\Omega^2$).

    \begin{figure}[!htb]\centering
        \vspace{3mm}
        \includegraphics[width=7.0in]{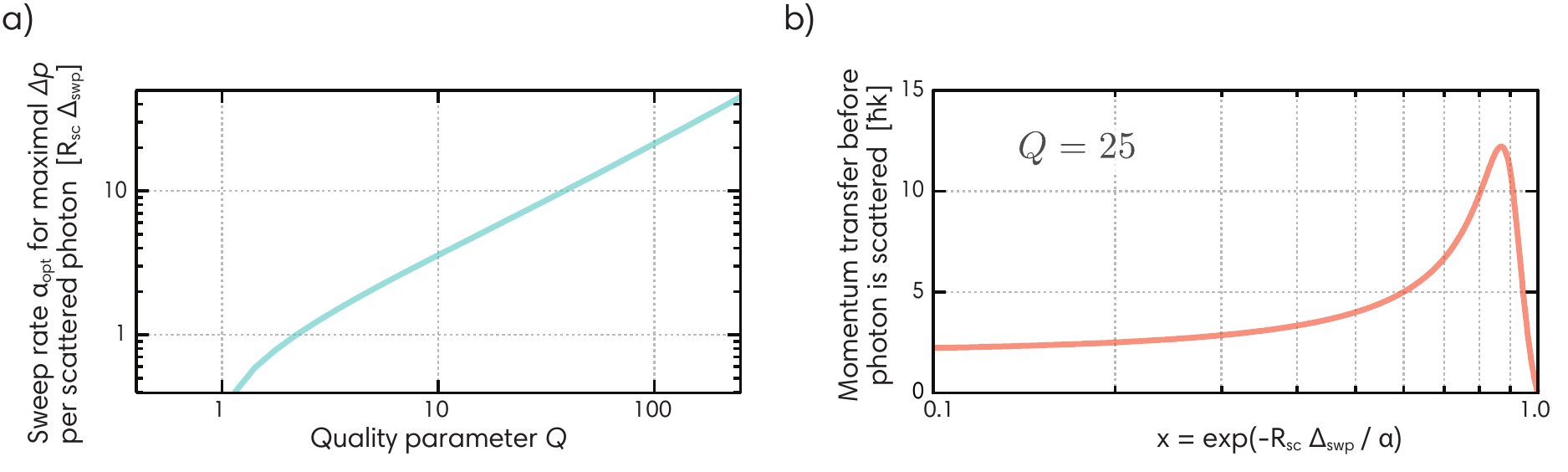}
        \caption{\textbf{(a)} The momentum transferred during a single simulated SWAP cooling sweep was found and maximized with respect to the sweep rate $\alpha$. When the quality parameter $Q$ is very small, the ideal strategy is to sweep very slowly in order to coherently scatter once. When $Q$ is large, efficient adiabatic sweeps are possible. \textbf{(b)} The maximum momentum transferable before a photon scatters due to Raman cooling beams or optical pumping beams.}
        \label{fig:momTransfer}
    \end{figure}



\end{spacing}
\end{widetext}
\end{document}